\documentclass[sn-mathphys,Numbered]{sn-jnl}

\usepackage{bm}
\usepackage{upgreek}
\usepackage{amsmath,lipsum}
\usepackage{graphicx}%
\usepackage{multirow}%
\usepackage{amsmath,amssymb,amsfonts}%
\usepackage{amsthm}%
\usepackage{mathrsfs}%
\usepackage[title]{appendix}%
\usepackage{xcolor}%
\usepackage{textcomp}%
\usepackage{manyfoot}%
\usepackage{booktabs}%
\usepackage{algorithm}%
\usepackage{algorithmicx}%
\usepackage{algpseudocode}%
\usepackage{listings}%
\usepackage{epstopdf}
\usepackage{subcaption}

\theoremstyle{thmstyleone}%
\theoremstyle{thmstyletwo}%

\theoremstyle{thmstylethree}%

\begin{document}

\title[Article Title]{Ultrasonic spin pumping in the antiferromagnetic acoustic resonator $\alpha-\text{Fe}_2\text{O}_3$}


\author*[1,2]{\sur{David A. Gabrielyan}}\email{davidgabrielyan1997@gmail.com}
\author[1,2]{\sur{Dmitry A. Volkov}}\email{d.a.volkov.work@gmail.com}
\equalcont{These authors contributed equally to this work.}
\author[1,3]{\sur{Tatyana V. Bogdanova}}\email{bogdanova.tv@phystech.edu}
\equalcont{These authors contributed equally to this work.}
\author[1,3]{\sur{Kristina D. Samoylenko}}\email{kris\_samoylenko@mail.ru}
\equalcont{These authors contributed equally to this work.}
\author[1,2]{\sur{Anton V. Matasov}}\email{matasov\_av93@mail.ru}
\equalcont{These authors contributed equally to this work.}
\author[1,2]{\sur{Ansar R. Safin}}\email{arsafin@gmail.com}
\equalcont{These authors contributed equally to this work.}
\author[1,3,4]{\sur{Dmitry V. Kalyabin}}\email{dmitry.kalyabin@phystech.edu}
\equalcont{These authors contributed equally to this work.}
\author[1,5]{\sur{Alexey A. Klimov}}\email{aleks-klimov@yandex.ru}
\equalcont{These authors contributed equally to this work.}
\author[6]{\sur{Leonid M. Krutyansky}}\email{leonid.krut@kapella.gpi.ru}
\equalcont{These authors contributed equally to this work.}
\author[6]{\sur{Vladimir L. Preobrazhensky}}\email{vlpreobr@yandex.ru}
\equalcont{These authors contributed equally to this work.}
\author[1,3,7]{\sur{Sergey A. Nikitov}}\email{nikitov@cplire.ru}
\equalcont{These authors contributed equally to this work.}

\affil*[1]{\orgname{Kotel'nikov Institute of Radioengineering and Electronics of Russian Academy of Sciences}, \orgaddress{\street{11 Mokhovaya}, \city{Moscow}, \postcode{125009}, \country{Russia}}}

\affil[2]{\orgname{National Research University "MPEI"}, \orgaddress{\street{17 Krasnokazarmennaya}, \city{Moscow}, \postcode{111250}, \country{Russia}}}

\affil[3]{\orgname{Moscow Institute of Physics and Technology}, \orgaddress{\street{9 Institutsky lane}, \city{Moscow}, \postcode{141701}, \state{Dolgoprudny}, \country{Russia}}}

\affil[4]{\orgname{HSE University}, \orgaddress{\street{20 Myasnitskaya}, \city{Moscow}, \postcode{101000}, \country{Russia}}}

\affil[5]{\orgname{MIREA – Russian Technological University}, \orgaddress{\street{78 Vernadsky Ave.}, \city{Moscow}, \postcode{119454}, \country{Russia}}}

\affil[6]{\orgname{Institute of General Physics named after. A.M. Prokhorov Russian Academy of Sciences}, \orgaddress{\street{38 Vavilova}, \city{Moscow}, \postcode{119991}, \country{Russia}}}

\affil[7]{\orgname{Saratov State University, Laboratory "Magnetic Metamaterials"}, \orgaddress{\street{83 Astrakhanskay}, \city{Saratov}, \postcode{410012}, \country{Russia}}}


\abstract{
Recent advances in magnon spintronics have ignited interest in the interactions between the spin and elastic subsystems of magnetic materials. These interactions suggest a dynamic connection between collective excitations of spins, quantized as magnons, and elastic waves generated by perturbations in the crystal lattice, quantized as phonons. Both magnons and their associated magnon-phonon excitations can act as sources of spin pumping from magnetic materials into non-magnetic metals. Although a considerable body of research has focused on spin pumping via elastic waves in ferromagnets, similar investigations involving antiferromagnets have yet to be undertaken. In this work, we experimentally demonstrate for the first time the feasibility of generating spin currents at ultrasonic frequencies of acoustic resonance in antiferromagnetic crystal hematite $\alpha-\text{Fe}_2\text{O}_3$ at room temperature. We provide both theoretical and experimental evidence that, due to strong magnetoelastic coupling, acoustic vibrations in hematite induce significant variable deviations in magnetization, resulting in spin accumulation at the antiferromagnet-normal metal interface, which in turn leads to the generation of spin and charge currents in the metal. Charge currents arising from the inverse spin Hall effect can be measured using the same methodology employed under high-frequency spin pumping conditions at the resonances of the magnetic subsystem itself. Moreover, the acoustic resonance in hematite is significantly more pronounced (by hundreds or even thousands of times) than in other quasiferromagnetic or antiferromagnetic systems, enabling the attainment of extremely large amplitudes of magnetic oscillations for spin pumping. This research highlights the new approach of utilizing acoustic spin pumping to manipulate spin currents in magnetic materials, particularly antiferromagnets. Our results expand the understanding of the potential for generating and detecting spin currents through acoustic spin pumping in magnetic materials and open new prospects for the development of reconfigurable, portable, and highly sensitive functional devices based on antiferromagnets across a wide range of frequencies at room temperature.}




\maketitle

\section{Introduction}\label{sec1}

In modern electronics, the possibility of creating new electronic devices is being explored through the application of spintronics and magnonics \cite{RevModPhys.90.015005}. Within these fields, the processes of spin or magnetic moment transfer in structures containing magnetic materials are being studied. Furthermore, there is renewed interest in studying magnetoelastic effects in magnetics, which are characterized by the significant influence of magnetoelastic interaction on the magnetic and elastic subsystems of the crystal \cite{Borovik1965, savchenko1964soviet, Seavey1972AcousticRI, ozhogin1972easy, dikshtein1974effect, ozhogin1977effective, gulyaev1997magnetoacoustic}. This influence manifests itself particularly in the high efficiency of magnetoelastic coupling via magnetostrictive deformations, as well as in the strong dependence of both linear and nonlinear elastic moduli on the intensity of the magnetic field. \cite{ozhogin1977effective}. When there is strong magnetoelastic coupling, it becomes possible to utilize the physical features and advantages of both subsystems: the spin excitations in magnetics and low intrinsic losses in the elastic subsystem at ultrasonic frequencies. Relatively low acoustic losses ensure higher quality factors of magnetoacoustic resonances over a wide frequency range from hundreds of kHz to hundreds of MHz, which significantly expands the observation area for spin dynamics effects compared to the natural range of intrinsic frequencies (from units to hundreds of GHz) of magnon devices \cite{strugatsky2007acoustic, fetisov2006bistability, ozhogin1988anharmonicity}.

In spintronics and magnonics, particular attention is currently being paid to multi-sublattice systems, such as antiferromagnets (AFM), since these materials, under certain conditions, exhibit an anomalously strong phonon-magnon coupling, which presents both fundamental and applied interest, for example, in spin pumping processes (SP) \cite{saitoh2006conversion,azevedo2005dc, PhysRevLett.107.066604, xu2016handbook}. Spin pumping experiments have been conducted with various AFM materials such as $\text{Cr}_2\text{O}_3$ \cite{li2020spin}, $\text{MnF}_2$ \cite{vaidya2020subterahertz, ross2015antiferromagentic}, and NiO \cite{stremoukhov2024strongly}. However, these materials have resonance frequencies ranging from 100 GHz to units of terahertz, which significantly complicates experimental studies of spin current effects. At the same time, there are AFM materials with weak ferromagnetism due to the Dzyaloshinskii-Moriya (DMI) interaction, leading to an uncompensated magnetic moment, making such AFM materials more susceptible to external magnetic fields \cite{moriya1960new, dzyaloshinsky1958thermodynamic}. The resonance frequencies of such materials contain not only antiferromagnetic modes in the spectrum but also quasi-ferromagnetic ones in the range of tens of gigahertz, allowing for experimental studies using standard microwave techniques.

Of particular interest at present is the antiferromagnets, which retains antiferromagnetic ordering at room temperature. Hematite is one of the most common magnetic materials in nature and one of the first AFM materials whose magnetic properties were experimentally studied \cite{morin1951electrical,
PhysRev.8.721}. The crystal structure of hematite belongs to the rhombohedral system with the space symmetry group $\text{D}^{6}_{3d}$ and with the Morin transition temperature $T_{M}$=260 K \cite{morin1951electrical}, below which it demonstrates an easy-axis antiferromagnetic ordering. At temperatures above $T_{M}$, spin reorientation occurs, and up to the Néel temperature $T_{N}$=948 K, a state with easy-plane anisotropy is retained. This AFM material is an excellent candidate for magnetoacoustic studies, as it is characterized by strong dynamic magnetoelastic coupling and anomalously high magnetoacoustic nonlinearity \cite{Seavey1972AcousticRI, dikshtein1974effect, ozhogin1977effective, gulyaev1997magnetoacoustic, ozhogin1991nonlinear}. Various functional devices based on AFM have previously been described: oscillators \cite{khymyn2017antiferromagnetic, sulymenko2017terahertz}, detectors \cite{khymyn2017antiferromagnetic_spin, gomonay2018narrow}, emitters and amplifiers \cite{khymyn2016transformation}, neuromorphic processors \cite{bradley2023artificial,sulymenko2018ultra}, memory elements \cite{kosub2017purely,fina2020flexible}, spectrum analyzers \cite{10.1063/1.5140552}, etc. Most of the proposed antiferromagnetic spintronics devices use antiferromagnet-normal metal heterostructures, where spin accumulation can occur at the boundary under high-frequency spin pumping, while in the normal metal layer, spin and electric currents are separated due to the inverse spin Hall effect. The spin current is most commonly detected indirectly by measuring the voltage induced by the inverse spin Hall effect at the opposite edges of the normal metal layer of the sample. Resonant spin pumping was experimentally studied on the quasi-ferromagnetic mode of hematite \cite{boventer2021room, wang2021spin,lebrun2020long, gabrielyan2024room} and iron borate with similar parameters \cite{gabrielyan2024microwave}. The quality factor of such resonances are tens at frequencies of tens of gigahertz and limits the real practical application of such crystals in electronics. At the same time, in addition to magnetic resonances, magnetoacoustic oscillations can be effectively excited in hematite crystals \cite{ozhogin1988anharmonicity, ozhogin1991nonlinear} at frequencies in the megahertz range with much higher quality factors reaching units and tens of thousands. The presence of strong magnetoelastic coupling makes it possible to tune the resonance frequency over a wide range using an external constant magnetic field. Despite the existence of several works on acoustic spin pumping \cite{Voskanian1967, fetisov2006bistability}, all use ferrogarnets (mainly yttrium iron garnet) as magnetic materials, while the possibilities of using antiferromagnets for acoustic spin pumping have not been previously studied.

The aim of this work is an experimental and theoretical study of acoustic spin pumping from a single crystal antiferromagnetic hematite $\alpha-\text{Fe}_2\text{O}_3$ into a normal metal layer at room temperature. We show both theoretically and experimentally that, due to the strong magnetoelastic coupling in the crystal, acoustic oscillations cause significant variable deviations in magnetization, which create spin accumulation at the antiferromagnet-normal metal interface, leading to the emergence of spin and charge currents, which we measure due to the inverse spin Hall effect (ISHE).

\section{Experimental setup}\label{sec2}

\begin{figure}[h]
\centering
\begin{subfigure}{.52\textwidth}
  \centering
  \includegraphics[width=1\linewidth]{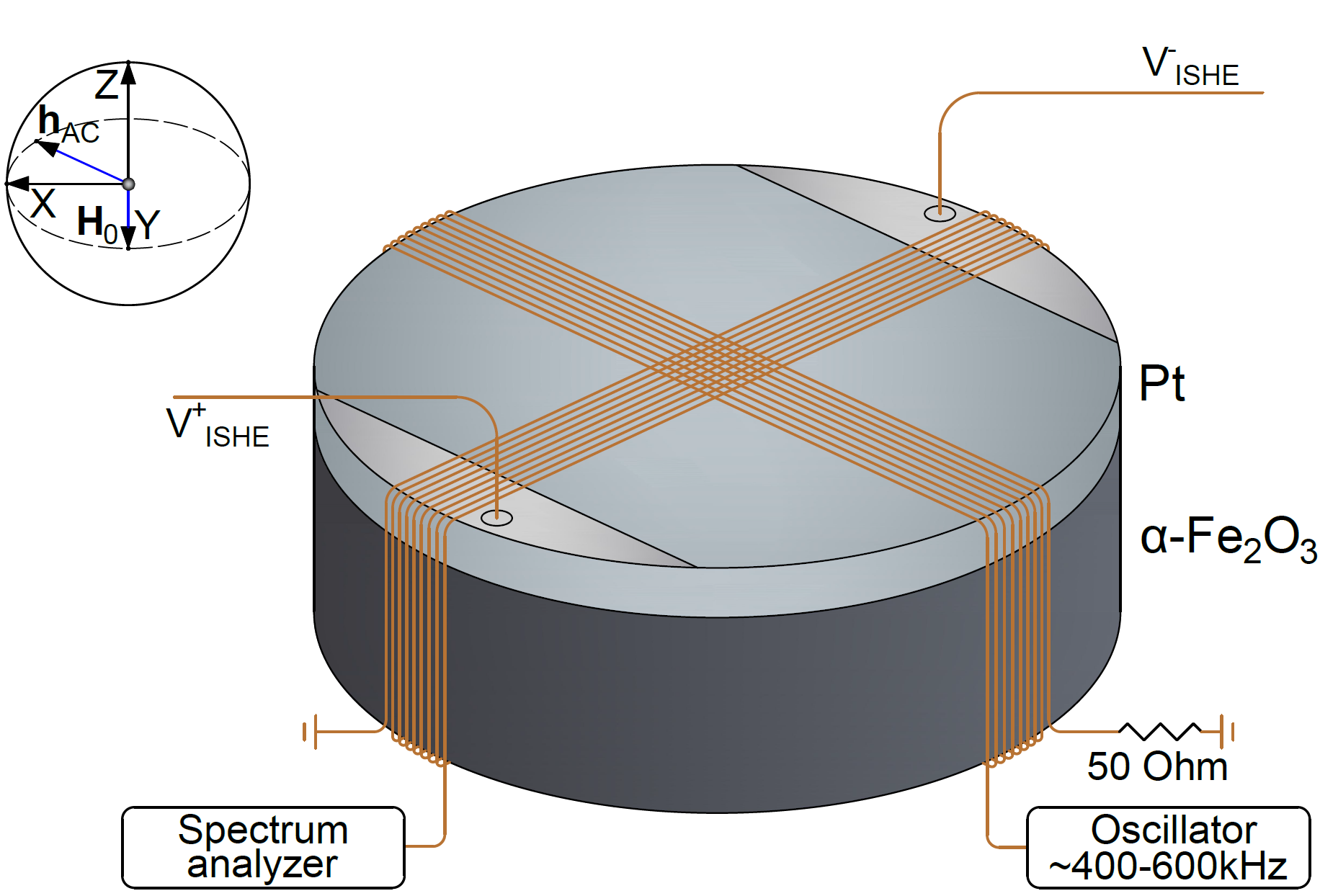}
  \caption{}
\end{subfigure}%
\begin{subfigure}{.52\textwidth}
  \centering
  \includegraphics[width=1\linewidth]{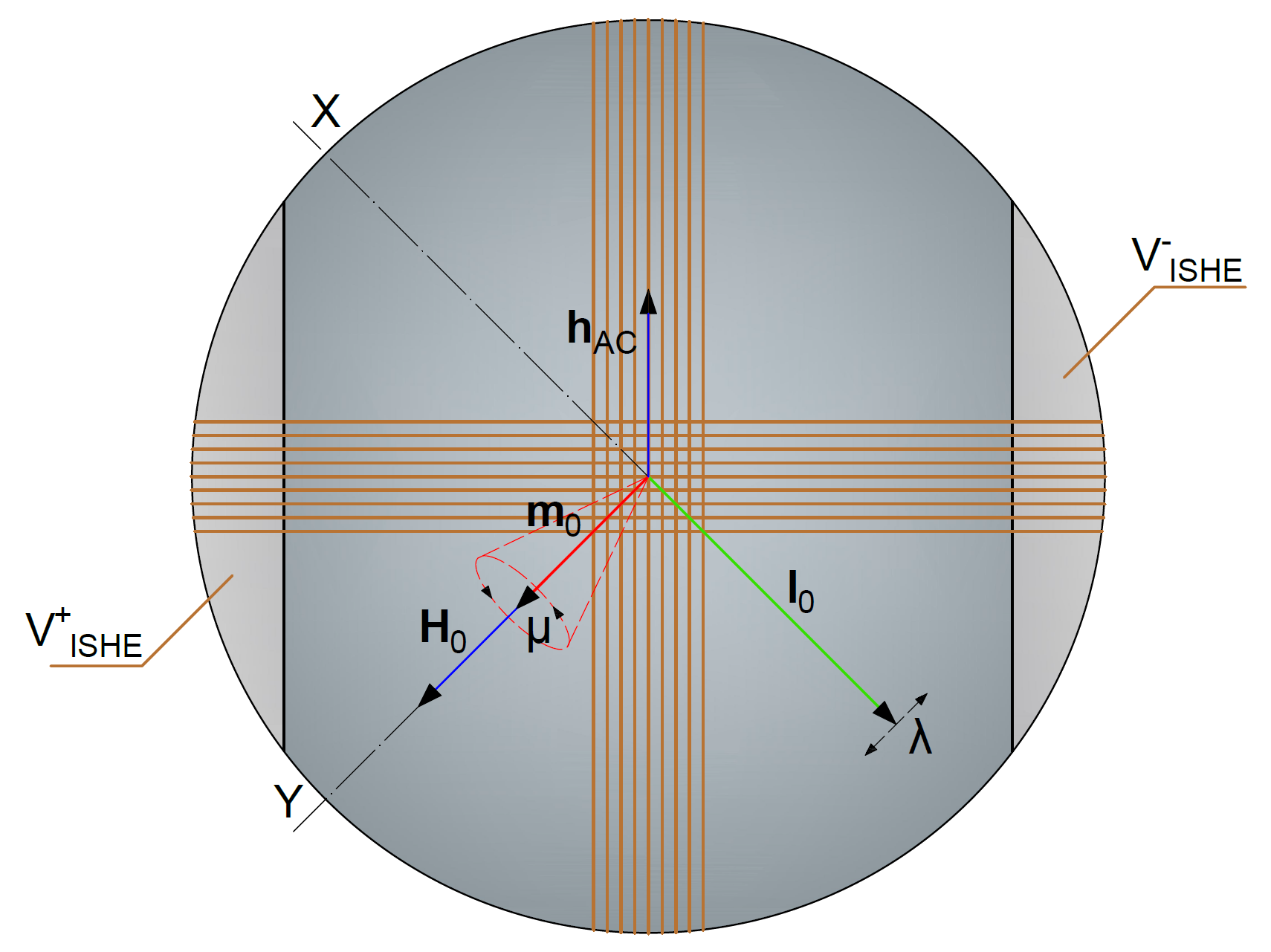}
  \caption{}
\end{subfigure}
\caption{\label{fig:Fig1} (a)Experimental setup for detecting magnetoelastic resonance and ISHE voltage. (b) Schematic of spin pumping and dynamics of the magnetization vectors $\mathbf{m_0}$ and $\mathbf{l_0}$ in the $\alpha-\text{Fe}_2\text{O}_3$ system.}
\end{figure}

We have used a single crystal disk (diameter – 5.5 mm, thickness – 500 µm, $c$-axis perpendicular to the sample plane) of hematite $\alpha-\text{Fe}_2\text{O}_3$ to study spin pumping with ultrasound. The single crystal samples were obtained using the flux method. Detailed conditions for the synthesis of single crystals are given in \cite{Voskanian1967}. X-ray powder diffraction phase analysis (XRD) of the samples are given in Appendix \ref{secA1}. A thin platinum layer 10 nm thick was deposited onto the surface of the single-crystal disk of hematite using magnetron sputtering. To excite and receive magnetoelastic resonance, two inductive coils were placed between the poles of a planar field electromagnet. The coils were oriented perpendicularly to each other to compensate for the direct transmission signal. Thus, when a harmonic signal with an excitation frequency out of acoustic resonance was applied to the primary (excitation) coil, no signal was observed on the secondary (receiving) coil. The first coil was connected to a sinusoidal signal oscillator, and the second one to the spectrum analyzer (see Fig.\ref{fig:Fig1}(a)). The ultrasonic oscillations of the antiferromagnetic disk were excited by an alternating magnetic field created by the coil as a result of the influence of a signal from the oscillator. If the excitation frequency was close to the frequency of the magnetoelastic resonance mode of the sample, then, due to the action of the alternating field $\mathbf{h}_\text{ac}(t)$, normal to the constant field $\mathbf{H}_0$, oscillations were excited in the sample, leading to signal excitation in the receiving coil.

For the experimental study of spin pumping, we excited magnetoacoustic oscillations in the resonator using an alternating magnetic field $\mathbf{h}_\text{ac}(t)$, generated by an amplitude-modulated signal in the frequency range from 400 kHz to 600 kHz with a modulation frequency of 977 Hz. The signal was applied from a signal oscillator to the excitation coil. The amplitude modulation of the excitation signal was used to implement the lock-in technique. Variable elastic deformations induce oscillations of the magnetization vector $\mathbf{m} = (\mathbf{M}_{1}+\mathbf{M}_{2})/2M_{0}$ and the Néel vector $\mathbf{l} = (\mathbf{M}_{1}-\mathbf{M}_{2})/ 2M_{0}$, where $\mathbf{M}_{1,2}$ are the magnetizations of the sublattices and $M_{0}$ is the saturation magnetization (see Fig.\ref{fig:Fig1}(b)). As a result of the oscillations $\mathbf{m}(t)$ at the boundary of the hematite and the normal metal layer, there is spin accumulation and pumping into the platinum layer (see Fig.\ref{fig:Fig1}(a)). The uncompensated spin current in platinum, due to the inverse spin Hall effect, causes charge separation at the contacts of the conductors with platinum. These conductors were connected to the measurement input of the lock-in amplifier. Electrical contact with the platinum was made using a conductive adhesive. The lock-in amplifier was phase-locked to the modulation frequency of the signal oscillator. It should be noted that, unlike standard microwave experiments \cite{gabrielyan2024room, gabrielyan2024microwave} on spin pumping (at frequencies in the gigahertz range), in this case, the alternating field operates at the acoustic resonance frequency of the disk, which is in the hundreds of kilohertz range.

\section{Magnetoacoustic spin-pumping}\label{sec3}

The mechanism of spin current generation by acoustic pumping will be considered within the framework of the theory of magnetoacoustic dynamics of an easy plane antiferromagnet (AFEP), developed in \cite{ozhogin1977effective, ozhogin1988anharmonicity, ozhogin1991nonlinear} based on a closed system of equations of motion of the antiferromagnetism vector. 

The energy density of the AFEP magnetic subsystem includes the interactions of the intersublattice exchange interaction, the Dzyaloshinskii-Moriya interaction energy, the anisotropy energy with the effective fields, respectively, ${H}_\text{E}$, ${H}_\text{D}$, ${H}_\text{a}$, the energy of interactions of the magnetic moment with the external constant $\mathbf{H}_0$, variable $\mathbf{h}_\text{ac}$ fields, and the energy of inhomogeneous exchange, the total density of the system will take the form:

\begin{equation}
\label{eq:1}
F_\text{m}=2M_{0}[2H_\text{E}\mathbf {m}^2 - \mathbf{H}_\text{D}[\mathbf{m}\times \mathbf{l}]_{z}+\frac{1}{2}H_\text{a}{l}^{2}_{z}+ \frac{1}{2} {\alpha} M_0 (\mathbf{\nabla} \mathbf{l})^2 -({\mathbf{m}\mathbf{H}_0} ) -(\mathbf{m} \mathbf{h}_\text{ac})],
\end{equation}
where $M_0$ is saturation magnetization and $\alpha$ is inhomogeneous exchange constant.

Considering the smallness of the relativistic fields compared to the exchange field, the ferromagnetic moment is defined as a function of the antiferromagnetic Neel vector, we find an expression for $\bf m$:

\begin{equation}
\label{eq:2}
\mathbf{m}= \frac{1}{\gamma H_\text{E}}[[\rm{\dot{\mathbf{l}}} \times \mathbf{l} ]+\gamma[\mathbf{l} \times[\mathbf{H}_0+\mathbf{h}_\text{ac}]]\times \mathbf{l} +\mathbf{l} \times \gamma \mathbf{H}_\text{D}]
\end{equation}
where $\gamma$ is the gyromagnetic ratio, $\mathbf{H}_\text{D} \parallel \mathbf{z}$, $ \mathbf{H}_0 \parallel \mathbf{y}$. All calculations are performed using the projection $\mathbf{h}_\text{ac} \parallel \mathbf{x}$. 

The spin current generation mechanism in AFEP under low-frequency ultrasonic pumping can be similar to the high-frequency pumping mechanism in ferrimagnets and ferrites. This similarity arises from the high amplitudes of magnetization oscillations observed in AFEP under acoustic resonance conditions. In the AFEP/heavy metal (HM) heterostructure, oscillations of the ferro- and antiferromagnetism vectors ($\mathbf m$ and $\mathbf l$) (see Fig.\ref{fig:Fig1}(b)) generate a spin current component polarized in the direction of the magnetizing field ($ \mathbf{H}_0 \parallel \mathbf{y}$) in the plane:

\begin{equation}
\label{eq:3}
I_\text{s} = g_\text{r} < [\mathbf{m} \times \dot{\mathbf{m}} ] + [\mathbf{l} \times \dot{\mathbf{l}}] >_{y},
\end{equation}
where $g_\text{r}$ is the spin mixing constant.

The dynamics of the N’eel vector in the considered area of frequencies, small compared to the frequency of antiferromagnetic resonance of the submillimeter range, is reduced to its turns in the basic plane, while the ferromagnetic moment precessifies with the exit from the plane. Under these conditions, only the first, ferromagnetic, component contributes to the spinal current. By expressing the magnetic vectors as the sum of the equilibrium and dynamic components, $\mathbf{l} = \mathbf{l}_0 + \boldsymbol{\lambda}(t)$, $\mathbf{m} = \mathbf{m}_0 + \boldsymbol{\mu}(t)$, and restricting the analysis to a linear approximation in terms of oscillation amplitudes, we can conclude that the contribution to the spin current is made by the projections of the ferromagnetic moment, which are determined, according to equation (\ref{eq:2}), by a single variable component of the Néel vector, $\mathrm{\lambda_{y}(t)}$:

\begin{equation}
\label{eq:4}
\begin{aligned}
{\mu}_{x} &= \frac{H_{0}+H_\text{D}}{2H_\text{E}} {\lambda}_{y}, \quad
{\mu}_{z} = \frac{\gamma^{-1}}{2H_\text{E}} \frac{\partial {\lambda}_{y}}{\partial t}.
\end{aligned}
\end{equation}

As a result, the constant component of the spin current is expressed by a general AFEP relation that is independent of the method used to excite magnetic oscillations:
\begin{equation}
\label{eq:5}
I_\text{s} = g_\text{r}\frac{\gamma(H_{0}+H_\text{D})}{(2H_\text{E} \gamma)^{2}} <\dot \lambda_{y}^{2}-\lambda_{y} \ddot \lambda_{y}>,
\end{equation}
where the bracket means averaging over time and over the crystal surface. 

Under conditions of harmonic oscillations of magnetization with frequency $\omega$, the spin current is equal to:
\begin{equation}
\label{eq:6}
I_\text{s} = g_\text{r}\frac{\gamma(H_{0}+H_\text{D})\omega^{2}}{2(\gamma H_\text{E})^{2}} <\lambda_{y}^{2}>.
\end{equation}

When describing resonant acoustic effects, the elastic displacement vector can be adequately represented as an expansion in the normal modes $\mathbf{u}_n(\mathbf {r})$ of acoustic oscillations: $\mathbf{u}_{n}(\mathbf {r},t)=\sum_{n} A_{n}(t) \mathbf{u}_n(\mathbf {r})$. Under the conditions of single-mode excitation of magnetoelastic oscillations by a uniform transverse harmonic field $h_\text{ac}(t)=h_\text{ac}(\omega)\text{e}^{i\omega t}+c.c.$, the variable component of the Néel vector is given by (see Appendix \ref{secA2}):

\begin{equation}
\label{eq:7}
\lambda_{y}(t,\mathbf{r})=\chi_{n}(\omega,\mathbf{r})h_\text{ac}(\omega)\text{e}^{i\omega t}+c.c.,
\end{equation}
where 

\begin{equation}
\label{eq:8}
\chi_{n}(\omega,\mathbf{r})= \left(\frac{\gamma}{\omega_{S0}}\right)^{2} (H_0+H_\text{D}) \left[ \sigma_{n}(\mathbf{r})\frac{\zeta_{n}^{2}}{ \left(1-\omega^{2}/\Omega_{n}^{2}+i\omega Q_{n}^{-1}/\Omega_{n} \right)}+1\right],
\end{equation}
where $\zeta_{n}^{2}$ is the coefficient of magnetoelastic coupling of the acoustic mode "n", $\sigma_{n}(\mathbf{r})$ is the shape factor proportional to the distribution of deformations in the mode, $\Omega_{n}$ and $Q_{n}$ are the eigenfrequency and quality factor of the mode.

The frequency-field dependence of the spin current is described by the relation:
\begin{equation}
\label{eq:9}
I_\text{s} = g_\text{r} \frac{\gamma(H_0+H_\text{D})\omega^{2}}{(\gamma H_\text{E})^{2}}\overline{\left|\chi_{n}(\omega,\mathbf{r})\right|^{2}}\left|h_\text{ac}(\omega)\right|^{2},
\end{equation}
where the overline denotes averaging over the surface of the AFEP/metal interface. Note that, under the typical condition of strong magnetoelastic coupling for hematite $\zeta_{n}^{2}Q_{n} \gg 1$, the unity term in brackets of equation \ref{eq:8} can be neglected, and the frequency dependence of the spin current reproduces the squared shape of the acoustic resonance line.

\section{Results and Discussion}\label{sec12}

\begin{figure*}[t]
\centering
\includegraphics[width=.67\linewidth]{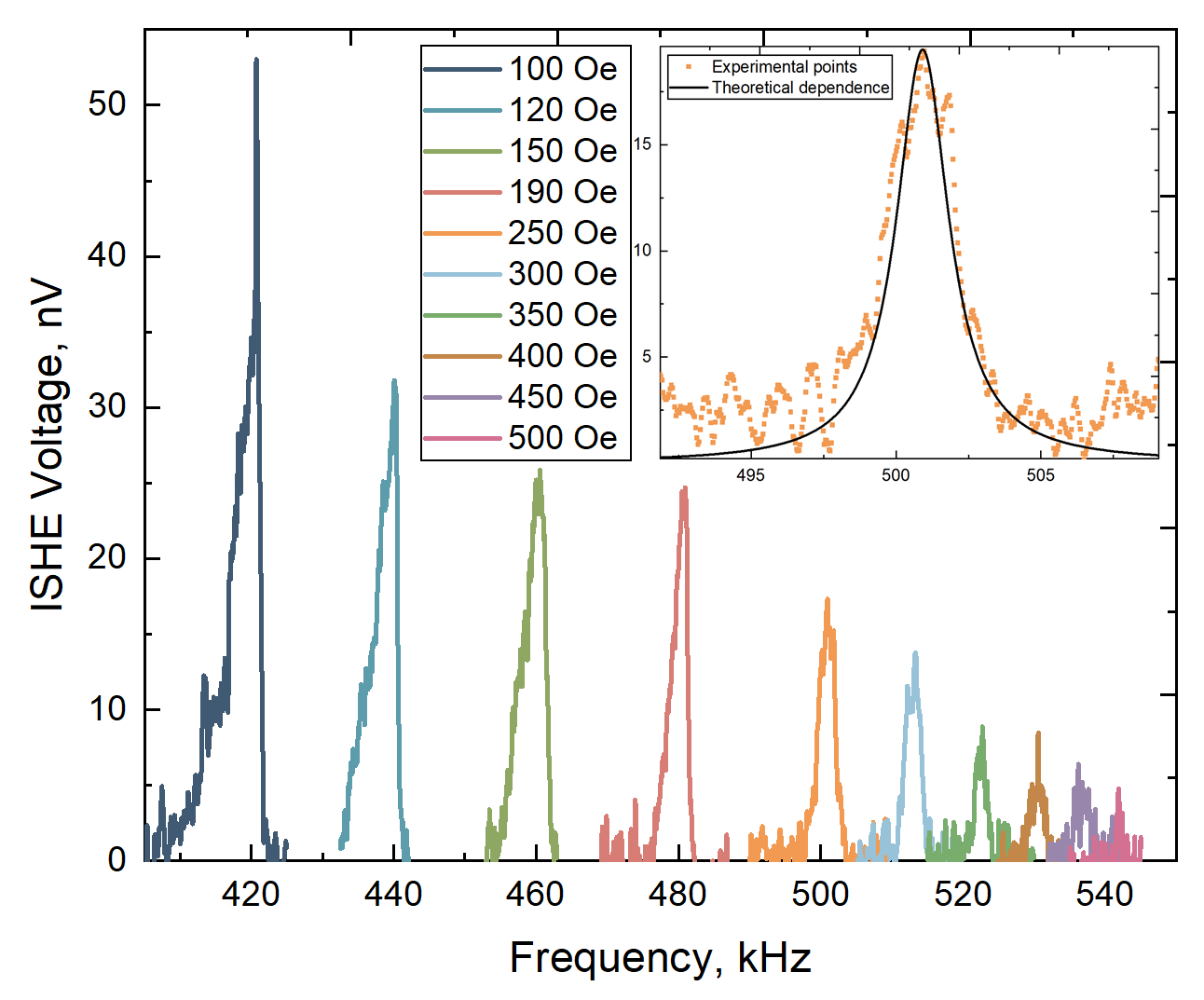}
\caption{\label{fig:Fig2}  Dependence of the ISHE voltage on frequency in the $\alpha-\text{Fe}_2\text{O}_3$/Pt disk at different values of the constant magnetic field. The inset shows the approximation of the ISHE voltage dependence by expression (\ref{eq:10}) at an external constant magnetic field value of 250 Oe.}
\end{figure*}

\begin{figure*}[h!]
\centering
\includegraphics[width=.67\linewidth]{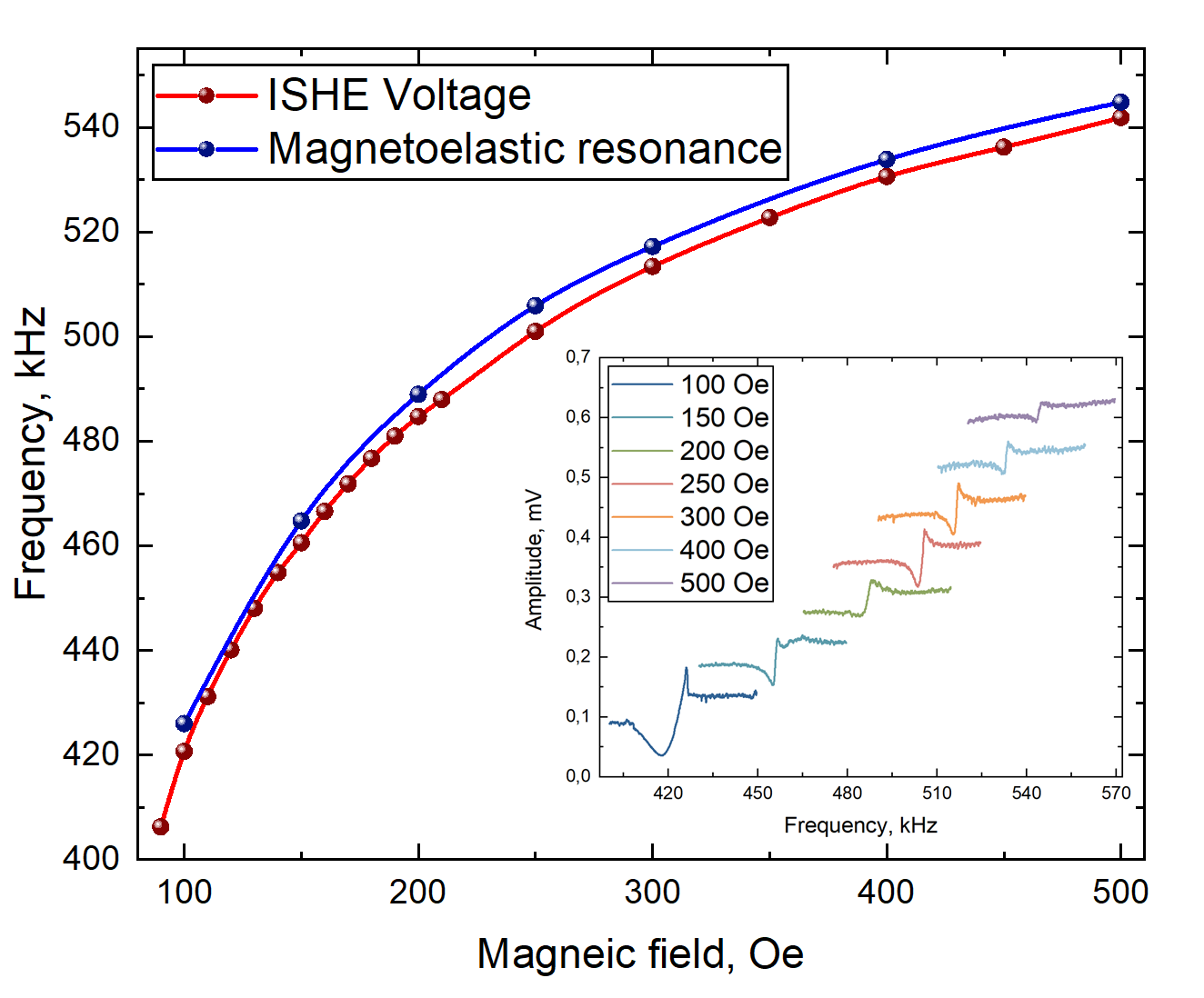}
\caption{\label{fig:Fig3} Dependences based on the values of resonant frequencies from the constant magnetic field for the frequency response of the resonator and the ISHE voltage. The insert shows the frequency response of the resonator, at different values of a constant magnetic field.}
\end{figure*}

We studied the $\alpha-\text{Fe}_2\text{O}_3$/Pt sample, where we experimentally demonstrate the efficient generation of the ISHE-rectified voltage using the setup shown in Fig.\ref{fig:Fig1}(a). Under the action of the excitation signal, the precession of the ferromagnetic vector is initiated indirectly, which leads to spin pumping at the $\alpha-\text{Fe}_2\text{O}_3$/Pt interface and injection of the spin current ($\mathbf{j_\text{sp}}$) into the platinum layer. Due to the inverse spin Hall effect, the resulting spin current in Pt is transformed into a charge current and into the ISHE Voltage. The measured ISHE voltages as a function of excitation frequency, obtained under different external constant magnetic fields, are shown in Fig.\ref{fig:Fig2}. The ISHE voltage is measured from points in the direction of the electric current, while the directions of the electric current, spin current, and polarization vector form a right-handed triple of vectors. This allows us to state that the ISHE voltage recorded during the experiment is proportional to the spin current from (\ref{eq:9}) \cite{10.1063/1.4977974}:

\begin{equation}
\label{eq:10}
V_\text{ISHE}=\Theta_\text{SH} \frac{\text{e} \lambda_\text{pt} \rho_\text{er} L}{2 \pi d_\text{pt}} \text{tanh}\left[\frac{d_\text{pt}}{2\lambda}\right] I_\text{s},
\end{equation}
where $\Theta_\text{SH}$ is the Hall spin angle, $\text{e}$ is the electron charge, $\lambda_\text{pt}$ is the spin diffusion length, $\rho_\text{er}$ is the specific electrical resistance, $L$ is the distance between the output electrodes, and $d_\text{pt}$ is the thickness of the platinum layer. The inset within Fig.\ref{fig:Fig2} shows the theoretical approximation of the experimentally obtained ISHE voltage at an external constant magnetic field of 250 Oe. Fig.$\ref{fig:Fig3}$ shows the dependence of the resonant frequencies on the constant magnetic field for the measurement of magnetoelastic resonance and ISHE voltage on the $\alpha-\text{Fe}_2\text{O}_3$/Pt heterostructure. A frequency shift is observed in the range from 100 to 500 kHz as the constant magnetic field changes. The alignment of the peaks of the magnetoelastic resonance and ISHE voltage confirms the correctness of the obtained data. The inset in Fig.\ref{fig:Fig3} shows the experimentally measured magnetoelastic resonance line of this structure at different values of the constant magnetic field.

\begin{figure*}[t!]
\centering
\includegraphics[width=.67\linewidth]{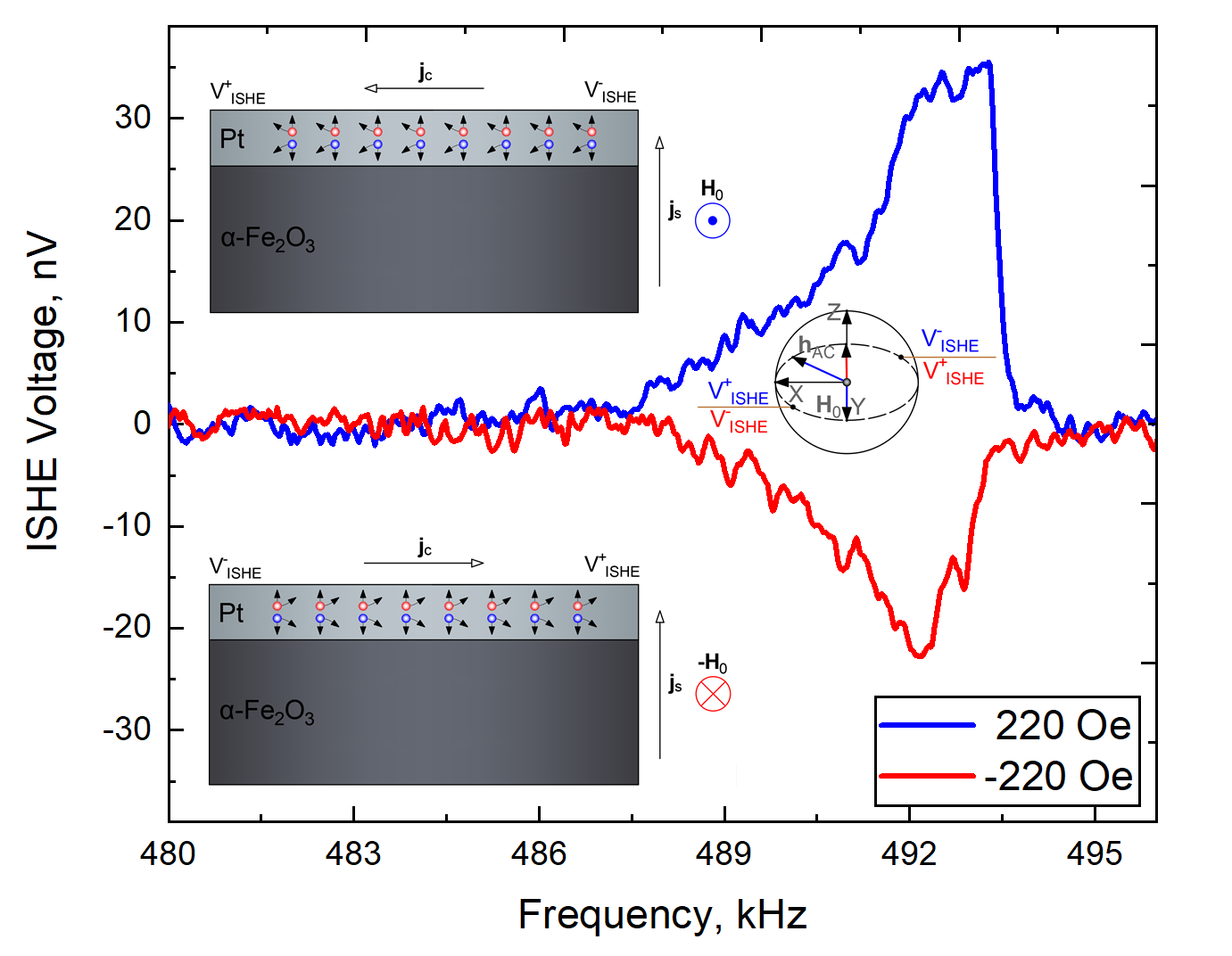}
\caption{\label{fig:Fig4} Experimental dependences of ISHE signals on the pumping frequency when the polarity of an external constant magnetic field with a strength of 220 Oe changes.}
\end{figure*}

A key feature of the ISHE mechanism is that the polarity of the spin current reverses when the direction of the magnetizing field is changed. This reversal is caused by the switch of sign of the Neel equilibrium vector. According to equation (\ref{eq:2}), the normal component of the ferromagnetic moment, $\mu_{0z}$, changes sign, while the tangential component, $\mu_{0x}$, remains unchanged. Consequently, the direction of polarization of the spin current (\ref{eq:3}) and, as a result, the direction of the charge current are inverted. To confirm that the measured voltage is caused by ISHE, an experiment was conducted to measure the ISHE voltage at different polarities of the magnetic field. Changing the direction of the external constant magnetic field to the opposite leads to a change in the direction of the main position of the vector $\bf (\textit{t})$ to the opposite, which, in turn, leads to a change in the sign of the voltage from the inverse spin Hall effect. Fig.\ref{fig:Fig4} illustrates the experimentally observed reversal of the electrical voltage at the measurement contacts when the magnetizing field polarity is switched. The shape of the resonance line is influenced by a nonlinear frequency shift characteristic of magnetoelastic oscillations in AFEP under high pumping conditions \cite{fetisov2006bistability}.

The difference in the resonance voltage modulus values, which can be seen in Fig.\ref{fig:Fig4}, is typical of such experiments and is associated with a number of side effects (see \cite{gabrielyan2024room, gabrielyan2024microwave} for more details).

\section{Conclusion}\label{sec13}

We have experimentally demonstrated the possibility of spin pumping at acoustic resonance frequencies from a disk of antiferromagnetic hematite $\alpha-\text{Fe}_2\text{O}_3$ by measuring the voltage caused by the inverse spin Hall effect in a normal metal layer. The obtained dependences are resonant in nature, and the characteristics "resonant frequency - magnetic field" constructed from them coincide with standard measurements using amplitude-frequency characteristics, which is consistent with theoretical results. We have shown theoretically and experimentally that, due to the strong magnetoelastic coupling in the material, acoustic oscillations cause deviations in the hematite magnetization, which create spin accumulation at the antiferromagnet-normal metal boundary, leading to the emergence of spin and charge currents that are measured due to the inverse spin Hall effect. Our results expand our understanding of the possibility of generating and detecting spin currents using acoustic spin pumping in magnetic materials and open up new prospects for the development of tunable, portable, and highly sensitive functional devices based on antiferromagnets in a wide frequency range at room temperatures.

\bmhead{Acknowledgements}
We thank Prof. Valery A. Murashov for the growth of $\alpha-\text{Fe}_2\text{O}_3$ single crystal. This work was carried out within the framework of the state assignment of the Kotelnikov Institute of Radioengineering and Electronics of the Russian Academy of Sciences.

\backmatter

\newpage
\bmhead{Supplementary information}

\begin{appendices}

\section{X-ray powder diffraction phase analysis}\label{secA1}

X-ray powder diffraction phase analysis (XRD) of the synthesized samples was conducted using an automated X-ray diffractometer, DRON 3, with filtered CoK$\alpha$ radiation. The phase composition of the studied samples was identified using the PDF2 database from the International Centre for Diffraction Data (ICDD)\cite{ICDD}. The unit cell parameters were calculated using CELREF software.

\begin{figure*}[h]
\centering
\includegraphics[width=.8\linewidth]{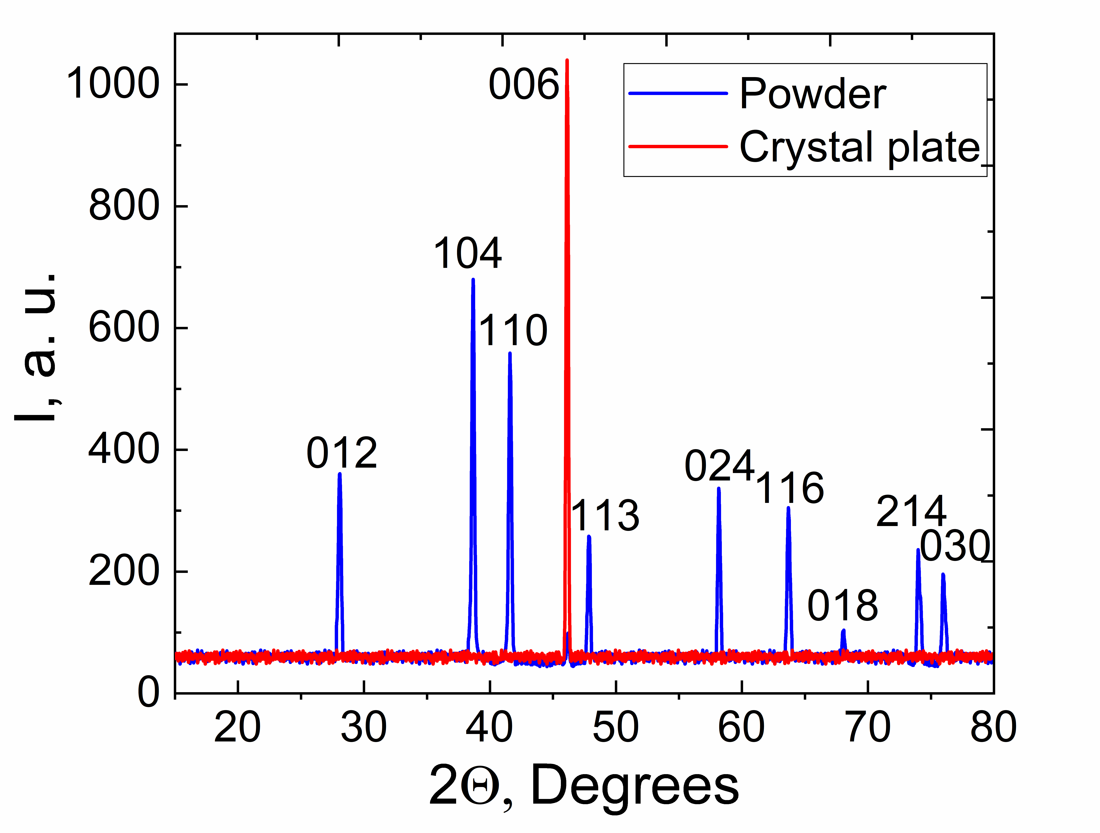}
\caption{\label{fig:FigA1} X-ray phase analysis of synthesized samples of $\alpha-\text{Fe}_2\text{O}_3$ using an automated X-ray diffractometer (DRON-3), where the X-ray peaks obtained from a single-crystal sample of $\alpha-\text{Fe}_2\text{O}_3$ (crystal plate) are shown in red, and the X-ray peaks from the powder obtained from $\alpha-\text{Fe}_2\text{O}_3$ (powder) are shown in blue.}
\end{figure*}

It is known that when synthesizing using the solution-melt method, you give in the form of thin plates, where the C axis is oriented to the support plane of the sample (Fig. \ref{fig:FigA1}).

All diffraction peaks (Fig. \ref{fig:FigA1}) of the crystalline powder are well indexed according to the hexagonal unit cell with the lattice parameters a = 5.0354(2) $\text{\r{A}}$, c = 13.7337(4) $\text{\r{A}}$. The orientation of single crystal plates along the c-axis was confirmed.

\section{Magnetoacoustic resonance model}\label{secA2}

The mechanism of spin current generation by acoustic pumping is considered within the framework of the theory of magnetoacoustic dynamics AFEP, developed in \cite{ozhogin1977effective, ozhogin1988anharmonicity} on the basis of a closed system of equations of motion of the antiferromagnetism vector. In the range of frequencies small compared to the frequency of the activation branch of the spin wave spectrum lying in the submillimeter wavelength range, the oscillations of the antiferromagnetic vector are described by the equation \cite{ozhogin1977effective}:

\begin{equation}
\label{eq:B1}
\gamma^{-2} \big[\mathbf{l} \times \ddot{\mathbf{l}} \big]_z = (\mathbf{H}_0 \cdot \mathbf{l} ) \big([\mathbf{H}_0 \times \mathbf{l}\big]_z + H_\text{D})  +  2 H_\text{E} M_{0} \alpha\left[\mathbf{l} \times \nabla^2 \mathbf{l}\right]_{z} + 2 H_\text{E} \left[\mathbf{l} \times \mathbf{H}_\text{me}\right]_{z},
\end{equation}
where the effective field of magnetoelastic interaction $\mathbf{H}_\text{me} = -\frac{1}{2 M_0} \frac{\partial F_\text{me}}{\partial \mathbf{l}}$. Taking into account the relative smallness of the ferromagnetic moment, in the magnetoelastic energy $F_\text{me}$ it is sufficient to take into account only the antiferromagnetic contribution $F_\text{me}=\mathbf{l}(\hat{B}\hat{u})\mathbf{l}$, where $\hat{u}$ is the elastic deformation tensor, $\hat{B}$ is the magnetostriction constant tensor. When describing the magnetoelastic dynamics of AFEP, it is convenient to move to angular variables, taking into account the conservation of the vector modulus $|\mathbf{l}|\approx 1$ in the exchange approximation $l_x=-\text{cos}{\varphi}$, $l_y=\text{sin}{\varphi}$. In this case, the magnetoelastic energy is transformed to the form:
\begin{equation}
\label{eq:B2}
F_\text{me}= \hat{B}_{1}\hat{u} \ \text{cos}2\varphi + \hat{B}_{2}\hat{u} \ \text{sin} 2\varphi.
\end{equation}

Elastic deformations $ \hat{{u}} = \hat{{u}}_{0}+\hat{{u}}(t)$ contain equilibrium $ \hat{{u}}_{0}$ and variable (acoustic) $\hat{u}(t)$ components. The equilibrium deformation is caused by spontaneous magnetostriction and makes a significant contribution to the activation of the quasi-ferromagnetic branch of the spin wave spectrum \cite{Borovik1965}. The linearized equation of motion (\ref{eq:B1}) in angular variables takes the form:

\begin{equation}
\label{eq:B3}
-\gamma^{-2} \left(\ddot{{\varphi}} - v_{S}^2\Delta {\varphi} \right)= {\gamma ^{ - 2}}\omega _{S0}^2\varphi  + \frac{{2{H_E}}}{{{M_0}}}{\hat B_2}\hat u(t) - {h_\text{ac}}(t)\left( {H + {H_D}} \right)
\end{equation}
when $ v_{S}$ is spin wave velocity, $h_\text{ac}$ is the component of the alternating field that excites magnetoelastic oscillations, $\omega_{S0}^2=\gamma^2 (H_0H_\text{D}+H_0^2+2H_\text{E}H_\text{me})$ is square of the resonance frequency of the quasi-ferromagnetic mode of the spin wave spectrum, $H_\text{me}=-\frac{2}{M_{0}}\hat{B}_{1}\hat{u}_{0}$ is the effective field of spontaneous magnetostriction.

In the long-wavelength approximation at frequencies $\omega \ll \omega_{S0}$, the solution of equation \ref{eq:B3} takes the form:

\begin{equation}
\label{eq:B4}
\mathbf{\varphi} = \left(\frac{\gamma}{\omega_{S0}}\right)^{2}\left[ -\frac{2 H_\text{E}}{M_{0}}\hat{B}_{2}\hat{u}(t) + h_\text{ac}(t)({H + {H_\text{D}}})\right]
\end{equation}

To describe resonant acoustic effects, we use the expansion of elastic displacements in normal modes $\mathbf{u}_{n}(r,t)$ of acoustic oscillations. The amplitudes of normal modes are described by the equation of oscillations:

\begin{equation}
\label{eq:B5}
m_{n}\left(\ddot{A}_{n} +2\delta \dot{A}_{n}\right) =-\frac{\partial}{\partial A_{n}}\int F d \mathbf{r},
\end{equation}
where $m_{n}=\int d \mathbf{r} \rho \mathbf{u}_{n}^{2}$ is the effective mass and $\delta_{n}$ is the attenuation coefficient of the acoustic mode, the integration is carried out over the volume of the crystal, while in the right-hand side of the expression the terms of the elastic energy and magnetoelastic subsystems $F=F_\text{e}+F_\text{me}$ are taken into account, where $F_\text{e} = \frac{1}{2} \hat{C} \hat{u} \hat{u}$. In the long-wave approximation, the renormalization of elastic moduli, according to equations \ref{eq:B2} and \ref{eq:B4}, takes the form:

\begin{equation}
\label{eq:B6}
\hat{C}^\text{eff}(H)=\hat{C}-\left(\frac{4H_\text{E}}{M_{0}}\right)\frac{1}{(\omega_{S0}/\gamma)^{2}}[\hat{B_{2}}]^{2}.
\end{equation}

Then the natural frequency of the mode $\Omega_{n}$ and the driving force $f_{n}(t)$ are determined according to the relation (Eq.\ref{eq:B5}):

\begin{equation}
\label{eq:B7}
\Omega_{n}^{2}(H)=\frac{1}{m_{n}}\int d\mathbf{r} \hat{C}(H)\hat{u}_{n}^{2}(\mathbf{r}),
\end{equation}

\begin{equation}
\label{eq:B8}
f_{n}(t)=-h_\text{ac}(t)\frac{H_\text{D}}{m_{n}(\omega_{S0}/\gamma)^{2}}\int d\mathbf{r} \left[\hat{B}_{2}\hat{u}_{n}(\mathbf{r})\right].
\end{equation}

Under conditions of linear single-mode excitation of oscillations by harmonic pumping, we represent the functions in the form $h_\text{ac}(\omega, t)=h(\omega)\text{e}^{i\omega t}+c.c.$, ${A_{n}}(\omega,t)={A_{n}}(\omega)\text{e}^{i\omega t}+c.c.$. Taking this form in expressions Eq.\ref{eq:B5}, we obtain ${A_{n}}(\omega,t)=h_\text{ac}\frac{H_\text{D}}{2\Delta \omega (\omega_{S0}/\gamma)^{2}}\int d\mathbf{r} \hat{B}_{2}\hat{u}_{n}(\mathbf{r})/ \rho \int d\mathbf{r} (\hat{u}_{n})^{2}$. Then the angular deviation (Eq.\ref{eq:B5}) is proportional to the applied alternating field:

\begin{equation}
\label{eq:B9}
{\varphi}(t,\mathbf{r})=\chi_{n}(\omega,\mathbf{r})h_\text{ac}(\omega)\text{e}^{i\omega t}+c.c.,
\end{equation}
where $\chi_{n}(\omega,\mathbf{r})= \left(\frac{\gamma}{\omega_{S0}}\right)^{2} H_\text{D} \left[ \sigma_{n}(\mathbf{r})\frac{\zeta_{n}^{2}}{(1-\omega^{2}/\Omega_{n}^{2}+i\omega Q_{n}^{-1}/\Omega_{n})}+1\right]$ is the effective susceptibility, $Q_{n}=\Omega_{n}/2\delta_{n}$ is the acoustic quality factor of the mode, \mbox{$\zeta_{n}^{2}=\frac{H_\text{E}}{M_{0}} \frac{\left( \int  d\mathbf{r} \hat{B}_{2}\hat{u}_{n}(\mathbf{r}) \right)^{2}}{(\omega_{S0}/\gamma)^{2} V \int d\mathbf{r} \hat{C}(H)\hat{u}_{n}^{2}}$} is the coefficient of magnetoelastic coupling of the "n"-th acoustic mode, $\sigma_{n}(\mathbf{r})=V\hat{B}_{2}\hat{u}_{n}(\mathbf{r})/\int d\mathbf{r}\hat{B}_{2}\hat{u}_{n}(\mathbf{r})$ is the shape factor depending on the distribution of deformations over the volume of the resonator V. 

Under conditions of harmonic magnetoelastic oscillations with frequency \mbox{$\omega (\lambda_{y}\approx \varphi \ll 1)$}, the relation (Eq.\ref{eq:B6}) for the spin current is determined through the angular deviation of the antiferromagnetic vector \ref{eq:B9} and thus:
\begin{equation}
\label{eq:B10}
I_\text{s} = g_\text{r} \frac{\gamma(H_{0}+H_\text{D})\Omega_{n}^{2}}{(\gamma H_\text{E})^{2}}\overline{\left[\chi_{n}(\omega,\mathbf{r})\right]^{2}}h_\text{ac}^{2},
\end{equation}
where the bar denotes the average on the surface of the AFEP/HM interface. 

\begin{figure*}[h]
\centering
\includegraphics[width=.8\linewidth]{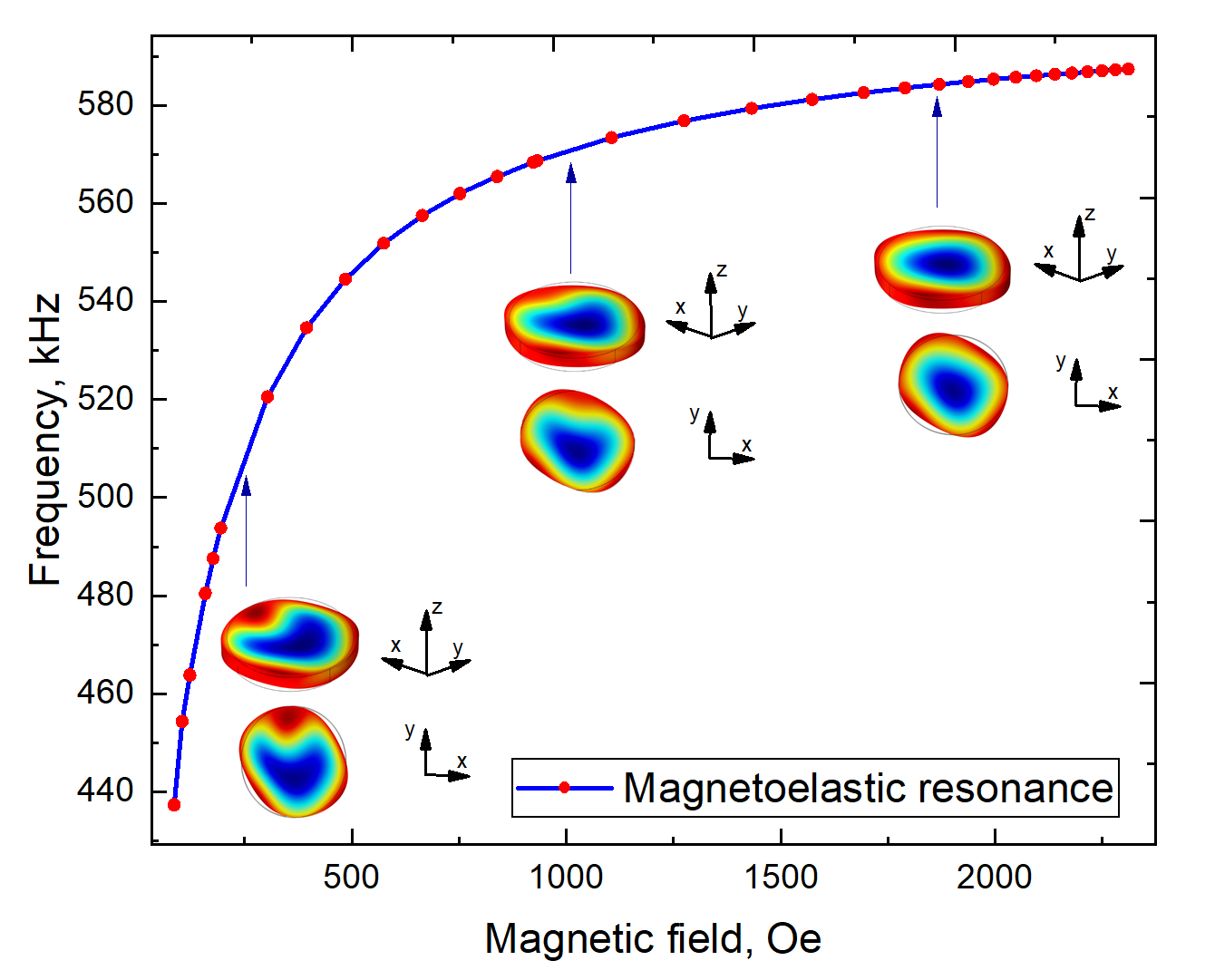}
\caption{\label{fig:FigB2} Dependence of the resonant frequency of the acoustic mode of samples of $\alpha-\text{Fe}_2\text{O}_3$ on the strength of the magnetizing field. The inset shows contour-shear modes excited in the $\alpha-\text{Fe}_2\text{O}_3$ resonator at an external magnetic field equal to 250, 1000 and 2000 Oe.}
\end{figure*}

We have obtained relations (Eq.\ref{eq:B10}), which determine the frequency-field dependences of the acoustically induced spin current. The natural frequencies of the EPAF resonators depend on the magnetizing field strength, while, according to (Eq.\ref{eq:B6} and Eq.\ref{eq:B7}), the field dependence is determined by the frequency of the quasi-ferromagnetic mode. The dependence of the frequency of the acoustic resonance on a relatively weak magnetic field (for hematite $H \ll H_\text{D}$=22 kOe) is approximated by the function \cite{preobrazhensky2010explosive}:

\begin{equation}
\label{eq:B11}
\Omega_{n}^{2}(H)=\Omega_{n0}^{2}(\infty)\left( 1 - \frac{H_{n}^{(1)}}{H+H_{n}^{(2)}} \right),
\end{equation}
where $\Omega_{n0}^{2}(\infty)$ is the maximum frequency corresponding to the saturation of magnetoacoustic coupling. 

For bulk magnetoelastic waves and modes with a strain distribution over the volume independent of the magnetic field, the effective field $H_{n}^{(2)}$ is equal to the spin-wave spectrum activation field $H_{n}^{(2)}=2H_\text{E}H_\text{me}/H_\text{D}$ \cite{Seavey1972AcousticRI}. In the modes of contour shear of thin plates, to which the mode used in these experiments belongs, the strains are subject to the requirement that the normal stresses on the plate surface vanish. As a result, the magnetoelastic field in $H_{n}^{(2)}$ is renormalized to $\overline{H_\text{me}}=H_\text{me}-\frac{2B_{14}^{2}}{M_{0}C_{44}}$. At the same time, the values of the fields $H_{n}^{(2)} \sim H_{n}^{(1)}$, which indicates a critical'softening' of the contour shear mode in the spin-reorientation instability region ($H \rightarrow 0$) \cite{ozhogin1991nonlinear, preobrazhensky2010explosive}.

To measure the parameters of the magnetoacoustic resonator, an extended range of magnetic field strength is used. Fig.\ref{fig:FigB2} shows the results of measuring the resonance frequency in the range from 50 Oe to 2.5 kOe. As a result of processing the obtained data, the resonator parameters were determined: $\Omega_{n}(H)/2\pi$ = 602 kHz, $H_{n}^{(2)}$ = 164 Oe, $H_{n}^{(1)}$ = 117 Oe. In Fig.\ref{fig:FigB2} we also present the acoustic mode structures and vibration spectra in the disk resonator in the paper. The results are provided by simulations using the COMSOL Multiphysics software \cite{multiphysics1998introduction} using the complet set of acoustic paramenters presented in \cite{moshkin2020wide}. Color coding indicates the deformation component $\mathbf{u}_{xy}$ from maximum values (red) to minimum values (blue) for the contour-shear mode. 

The parameters we used for the analytical calculations of $V_\text{ISHE}$ are shown in the form of Table \ref{tab3}.

\begin{table*}[t]
\caption{Numerical values of the parameters used to calculate the characteristics}\label{tab3}
\centering

\begin{tabular}{|c|c|c|}
\hline
Designation of a constant & Value of a constant & Dimension \\ \hline
    $\Theta_\text{SH}$   & 0.1 &  -- \\ \hline
    $\lambda_\text{Pt}$  & 7.3$\cdot$$10^{-9}$ & m \\ \hline
    $\rho$   & 4.8$\cdot$$10^{-7}$ & $\Omega \cdot$m \\ \hline
    $d_\text{Pt}$     & $10^{-8}$ &  m \\ \hline
    $g_\text{r}$     & 6.9$\cdot$$10^{18}$ & m$^{-2}$ \\ \hline
    $d_\text{he}$   & 5$\cdot$$10^{-4}$ & m \\ \hline
    $L$  & 5.5$\cdot$$10^{-3}$ & m \\ \hline
    $\gamma$ & 1.76$\cdot$10$^{7}$ & Hz$\cdot$Oe$^{-1}$ \\ \hline
    $H_\text{D}$     & 22$\cdot$$10^{-3}$ &  Oe \\ \hline
    $2H_\text{E}H_\text{me}$     &  4$\cdot$$10^{-6}$ & Oe$^{2}$ \\ \hline
    $h_\text{ac}$         & 3$\cdot$$10^{-2}$  &  Oe \\ \hline
    $\Omega_{n0}$ & 580$\cdot$$10^{3}$ & Hz \\ \hline
    $\delta$ & $10^{-3}$ & -- \\ \hline
    $M_0$         & 870$\cdot$$10^{3}$  &  emu/cm$^{3}$ \\ \hline
    $C_{44}$ & 0.85$\cdot$$10^{11}$ & Pa \\ \hline
    $2B_{14}$ & 27$\cdot$$10^{5}$ & Pa \\ \hline

\end{tabular}
\end{table*}

\end{appendices}


\newpage
\begin{thebibliography}{99}

\bibitem{RevModPhys.90.015005}
Baltz, V., Manchon, A., Tsoi, M., Moriyama, T., Ono, T., Tserkovnyak, Y.:
Antiferromagnetic spintronics.
\textit{Rev. Mod. Phys.} \textbf{90}(1), 015005 (2018)

\bibitem{Borovik1965}
Borovik-Romanov, A.S., Rudashevskii, E.G.:
Effect of spontaneous striction on antiferromagnetic resonance in hematite.
\textit{Soviet Phys. JETP} \textbf{20}, 1407--1411 (1965)

\bibitem{savchenko1964soviet}
Savchenko, M.A.:
Soviet Physics.
\textit{Solid State} \textbf{6}, 666 (1964)

\bibitem{Seavey1972AcousticRI}
Seavey, M.H.:
Acoustic resonance in the easy-plane weak ferromagnets $\alpha$-Fe$_2$O$_3$ and FeBO$_3$.
\textit{Solid State Communications} \textbf{10}, 219--223 (1972)

\bibitem{ozhogin1972easy}
Ozhogin, V., Maximenkov, P.:
Easy plane antiferromagnets (AFEP) for applications: Hematite.
\textit{IEEE Transactions on Magnetics} \textbf{8}(3), 645--645 (1972)

\bibitem{dikshtein1974effect}
Dikshtein, I.E., Tarasenko, V.V., Shavrov, V.G.:
Effect of pressure on magnetoacoustic resonance in uniaxial antiferromagnets.
\textit{Zh. Eksp. Teor. Fiz.} \textbf{67}(2), 816--823 (1974)

\bibitem{ozhogin1977effective}
Ozhogin, V.I., Preobrazhenskii, V.L.:
Effective anharmonicity of elastic subsystem of antiferromagnets.
\textit{Sov. Phys. JETP} \textbf{46}, 523--529 (1977)

\bibitem{gulyaev1997magnetoacoustic}
Gulyaev, Y.V., Dikshtein, I.E., Shavrov, V.G.:
Magnetoacoustic surface waves in magnetic crystals near spin-reorientation phase transitions.
\textit{Physics-Uspekhi} \textbf{40}(7), 701 (1997)

\bibitem{strugatsky2007acoustic}
Strugatsky, M.B., Skibinsky, K.M.:
Acoustic resonances in antiferromagnet FeBO$_3$.
\textit{J. Magn. Magn. Mater.} \textbf{309}(1), 64--70 (2007)

\bibitem{fetisov2006bistability}
Fetisov, Y.K., Preobrazhenskii, V.L., Pernod, P.:
Bistability in a nonlinear magnetoacoustic resonator.
\textit{J. Commun. Technol. Electron.} \textbf{51}, 218--230 (2006)

\bibitem{ozhogin1988anharmonicity}
Ozhogin, V.I., Preobrazhenskii, V.L.:
Anharmonicity of mixed modes and giant acoustic nonlinearity of antiferromagnetics.
\textit{Soviet Physics Uspekhi} \textbf{31}(8), 713--728 (1988)

\bibitem{saitoh2006conversion}
Saitoh, E., Ueda, M., Miyajima, H., Tatara, G.:
Conversion of spin current into charge current at room temperature: Inverse spin-Hall effect.
\textit{Appl. Phys. Lett.} \textbf{88}(18) (2006)

\bibitem{azevedo2005dc}
Azevedo, A., Leão, L.H.V., Rodriguez-Suarez, R.L., Oliveira, A.B., Rezende, S.M.:
dc effect in ferromagnetic resonance: Evidence of the spin-pumping effect?
\textit{J. Appl. Phys.} \textbf{97}(10) (2005)

\bibitem{PhysRevLett.107.066604}
Heinrich, B., Burrowes, C., Montoya, E., Kardasz, B., Girt, E., Song, Y.-Y., Sun, Y., Wu, M.:
Spin Pumping at the Magnetic Insulator (YIG)/Normal Metal (Au) Interfaces.
\textit{Phys. Rev. Lett.} \textbf{107}(6), 066604 (2011)

\bibitem{xu2016handbook}
Xu, Y., Awschalom, D.D., Nitta, J.:
\textit{Handbook of Spintronics}.
Springer Publishing Company, New York (2016)

\bibitem{li2020spin}
Li, J., Wilson, C.B., Cheng, R., Lohmann, M., Kavand, M., Yuan, W., Aldosary, M., Agladze, N., Wei, P., Sherwin, M.S., et al.:
Spin current from sub-terahertz-generated antiferromagnetic magnons.
\textit{Nature} \textbf{578}(7793), 70--74 (2020)

\bibitem{vaidya2020subterahertz}
Vaidya, P., Morley, S.A., van Tol, J., Liu, Y., Cheng, R., Brataas, A., Lederman, D., Del Barco, E.:
Subterahertz spin pumping from an insulating antiferromagnet.
\textit{Science} \textbf{368}(6487), 160--165 (2020)

\bibitem{ross2015antiferromagentic}
Ross, P., Schreier, M., Lotze, J., Huebl, H., Gross, R., Goennenwein, S.T.B.:
Antiferromagentic resonance detected by direct current voltages in MnF$_2$/Pt bilayers.
\textit{J. Appl. Phys.} \textbf{118}(23) (2015)

\bibitem{stremoukhov2024strongly}
Stremoukhov, P., Safin, A., Schippers, C.F., Lavrijsen, R., Bal, M., Zeitler, U., Sadovnikov, A., Kozlova, E., Ilkhchy, K.S., Nikitov, S., et al.:
Strongly nonlinear antiferromagnetic dynamics in high magnetic fields.
\textit{Results in Physics} \textbf{57}, 107377 (2024)

\bibitem{moriya1960new}
Moriya, T.:
New mechanism of anisotropic superexchange interaction.
\textit{Phys. Rev. Lett.} \textbf{4}(5), 228 (1960)

\bibitem{dzyaloshinsky1958thermodynamic}
Dzyaloshinsky, I.:
A thermodynamic theory of “weak” ferromagnetism of antiferromagnetics.
\textit{J. Phys. Chem. Solids} \textbf{4}(4), 241--255 (1958)

\bibitem{morin1951electrical}
Morin, F.J.:
Electrical properties of $\alpha$-Fe$_2$O$_3$and $\alpha$-Fe$_2$O$_3$ containing titanium.
\textit{Phys. Rev.} \textbf{83}(5), 1005 (1951)

\bibitem{PhysRev.8.721}
Smith, T.T.:
The Magnetic Properties of Hematite.
\textit{Phys. Rev.} \textbf{8}(6), 721--737 (1916)

\bibitem{ozhogin1991nonlinear}
Ozhogin, V.I., Preobrazhenskii, V.L.:
Nonlinear dynamics of coupled systems near magnetic phase transitions of the “order-order” type.
\textit{J. Magn. Magn. Mater.} \textbf{100}(1--3), 544--571 (1991)

\bibitem{khymyn2017antiferromagnetic}
Khymyn, R., Lisenkov, I., Tiberkevich, V., Ivanov, B.A., Slavin, A.:
Antiferromagnetic THz-frequency Josephson-like oscillator driven by spin current.
\textit{Sci. Rep.} \textbf{7}, 43705 (2017)

\bibitem{sulymenko2017terahertz}
Sulymenko, O.R., Prokopenko, O.V., Tiberkevich, V.S., Slavin, A.N., Ivanov, B.A., Khymyn, R.S.:
Terahertz-frequency spin Hall auto-oscillator based on a canted antiferromagnet.
\textit{Phys. Rev. Appl.} \textbf{8}(6), 064007 (2017)

\bibitem{khymyn2017antiferromagnetic_spin}
Khymyn, R., Tiberkevich, V., Slavin, A.:
Antiferromagnetic spin current rectifier.
\textit{AIP Adv.} \textbf{7}(5), 055931 (2017)

\bibitem{gomonay2018narrow}
Gomonay, O., Jungwirth, T., Sinova, J.:
Narrow-band tunable terahertz detector in antiferromagnets via staggered-field and antidamping torques.
\textit{Phys. Rev. B} \textbf{98}(10), 104430 (2018)

\bibitem{khymyn2016transformation}
Khymyn, R., Lisenkov, I., Tiberkevich, V.S., Slavin, A.N., Ivanov, B.A.:
Transformation of spin current by antiferromagnetic insulators.
\textit{Phys. Rev. B} \textbf{93}(22), 224421 (2016)

\bibitem{bradley2023artificial}
Bradley, H., Louis, S., Trevillian, C., Quach, L., Bankowski, E., Slavin, A., Tyberkevych, V.:
Artificial neurons based on antiferromagnetic auto-oscillators as a platform for neuromorphic computing.
\textit{AIP Adv.} \textbf{13}(1) (2023)

\bibitem{sulymenko2018ultra}
Sulymenko, O., Prokopenko, O., Lisenkov, I., Åkerman, J., Tyberkevych, V., Slavin, A.N., Khymyn, R.:
Ultra-fast logic devices using artificial “neurons” based on antiferromagnetic pulse generators.
\textit{J. Appl. Phys.} \textbf{124}(15) (2018)

\bibitem{kosub2017purely}
Kosub, T., Kopte, M., Hühne, R., Appel, P., Shields, B., Maletinsky, P., Hübner, R., Liedke, M.O., Fassbender, J., Schmidt, O.G., et al.:
Purely antiferromagnetic magnetoelectric random access memory.
\textit{Nat. Commun.} \textbf{8}, 13985 (2017)

\bibitem{fina2020flexible}
Fina, I., Dix, N., Menendez, E., Crespi, A., Foerster, M., Aballe, L., Sanchez, F., Fontcuberta, J.:
Flexible antiferromagnetic FeRh tapes as memory elements.
\textit{ACS Appl. Mater. Interfaces} \textbf{12}(13), 15389--15395 (2020)

\bibitem{10.1063/1.5140552}
Artemchuk, P.Y., Sulymenko, O.R., Louis, S., Li, J., Khymyn, R.S., Bankowski, E., Meitzler, T., Tyberkevych, V.S., Slavin, A.N., Prokopenko, O.V.:
Terahertz frequency spectrum analysis with a nanoscale antiferromagnetic tunnel junction.
\textit{J. Appl. Phys.} \textbf{127}(6), 063905 (2020)

\bibitem{boventer2021room}
Boventer, I., Simensen, H.T., Anane, A., Kläui, M., Brataas, A., Lebrun, R.:
Room-temperature antiferromagnetic resonance and inverse spin-Hall voltage in canted antiferromagnets.
\textit{Phys. Rev. Lett.} \textbf{126}(18), 187201 (2021)

\bibitem{wang2021spin}
Wang, H., Xiao, Y., Guo, M., Lee-Wong, E., Yan, G.Q., Cheng, R., Du, C.R.:
Spin pumping of an easy-plane antiferromagnet enhanced by Dzyaloshinskii–Moriya interaction.
\textit{Phys. Rev. Lett.} \textbf{127}(11), 117202 (2021)

\bibitem{lebrun2020long}
Lebrun, R., Ross, A., Gomonay, O., Baltz, V., Ebels, U., Barra, A.-L., Qaiumzadeh, A., Brataas, A., Sinova, J., Kläui, M.:
Long-distance spin-transport across the Morin phase transition up to room temperature in ultra-low damping single crystals of the antiferromagnet $\alpha$-Fe$_2$O$_3$.
\textit{Nat. Commun.} \textbf{11}, 6332 (2020)

\bibitem{gabrielyan2024room}
Gabrielyan, D., Volkov, D., Kozlova, E., Safin, A., Kalyabin, D., Nikitov, S.:
Room-temperature spin pumping from canted antiferromagnet $\alpha$-Fe$_2$O$_3$.
\textit{J. Appl. Phys.} \textbf{136}(8) (2024)

\bibitem{gabrielyan2024microwave}
Gabrielyan, D.A., Volkov, D.A., Kozlova, E.E., Safin, A.R., Kalyabin, D.V., Klimov, A.A., Preobrazhensky, V.L., Strugatsky, M.B., Yagupov, S.V., Moskal, I.E., et al.:
Microwave spin-pumping from an antiferromagnet FeBO$_3$.
\textit{J. Phys. D: Appl. Phys.} \textbf{57}(30), 305003 (2024)

\bibitem{Voskanian1967}
Voskanian, R.L., Zheludev, I.S.:
Obtaining monocrystalline rhombohedra and plates of hematite [in Russian].
\textit{Crystallography} \textbf{12}(3), 539--540 (1967)

\bibitem{10.1063/1.4977974}
Khymyn, R., Tiberkevich, V., Slavin, A.:
Antiferromagnetic spin current rectifier.
\textit{AIP Adv.} \textbf{7}(5), 055931 (2017)

\bibitem{ICDD}
ICDD, PDF-2, Software version 4.19.21, database version 2.1901.
\url{http://www.icdd.com/pdfsearch/} (2019)

\bibitem{preobrazhensky2010explosive}
Preobrazhensky, V., Yevstafyev, O., Pernod, P., Berzhansky, V.:
Explosive instability of quasi-phonon triads in antiferromagnet under frequency modulated electromagnetic field.
\textit{J. Magn. Magn. Mater.} \textbf{322}(6), 585--588 (2010)

\bibitem{multiphysics1998introduction}
COMSOL Multiphysics:
Introduction to COMSOL Multiphysics\textsuperscript{\textregistered}.
\textit{COMSOL Multiphysics}, Burlington, MA, \textbf{9}, accessed Feb (1998)

\bibitem{moshkin2020wide}
Moshkin, V., Preobrazhensky, V., Pernod, P.:
Wide-range frequency control in magnetoacoustic resonator.
\textit{IEEE Trans. Ultrason. Ferroelectr. Freq. Control} \textbf{67}(9), 1957--1959 (2020)

\end{thebibliography}
\end{document}